\documentclass{elsart1p}
\usepackage{graphicx}
\usepackage{amssymb}

\newcommand{\be}{\begin{eqnarray}}

\newcommand{\beq}{\begin{equation}}
\newcommand{\eeq}{\end{equation}}

\newcommand{\bea}{\begin{eqnarray}}
\newcommand{\eea}{\end{eqnarray}}
\newcommand{\beqa}{\begin{eqnarray}}
\newcommand{\eeqa}{\end{eqnarray}}
\newcommand{\ba}{\begin{array}}
\newcommand{\ea}{\end{array}}

\newcommand{\noi}{\noindent}

\newcommand{\Th}{$\Theta^+\,$}

\begin{document}

\begin{frontmatter}

\title{Justifying the exotic $\Theta^+$ pentaquark}
\author{Dmitri Diakonov}
\address{Petersburg Nuclear Physics Institute, Gatchina 188300, St. Petersburg, Russia}

\begin{abstract}
The existence of a light $S\!=\!+1$ baryon resonance follows from Quantum Field Theory
applied to baryons. This is illustrated in the Skyrme model (where \Th exists but is too strong)
and in a new mean field approach where \Th arises as a consequence of three known resonances:
$\Lambda(1405)$, $N(1440)$ and $N(1535)$.
\end{abstract}
\end{frontmatter}

After strong signals of the exotic $S\!=\!+1$ baryon have been announced
in the Fall of 2002 from two independent searches~\cite{Nakano-1,Dolgolenko-1}
initiated by a theoretical prediction of a {\em light and narrow} exotic resonance~\cite{DPP-97},
there has been much confusion in the field, both on the experimental and theoretical
sides. The anomalously small width is the key to that confusion. If
$\Gamma_\Theta\leq 1\,{\rm MeV}$~\cite{Dolgolenko-2} it means that the $(\Theta KN)$
coupling is an order of magnitude less than the typical pseudoscalar couplings of normal
mainly $3Q$ baryons, and one expects that the vector coupling $(\Theta K^*N)$ is similarly
strongly suppressed. The \Th production cross sections are then very small and it indeed
becomes a challenge to reveal it experimentally.

Theoretically, both the light mass and a small width are uncomprehensible from the
na\"ive point of view on hadrons, which assumes one adds up $\approx 350\,{\rm MeV}$ per constituent
quark, plus 150 MeV for strangeness. With that arithmetics, one expects a pentaquark
$uudd\bar s$ state at $\geq 1900\,{\rm MeV}$ (400 MeV heavier than in reality!) and,
of course, there is no way to avoid a large width from a fall-apart decay.

There have been strong warnings in the past that the constituent quark picture is an
oversimplification, the ``spin crisis'' and the anomalously huge nucleon $\sigma$-term
being examples of such warnings~\cite{D-04}. However it is the light and narrow \Th
that is really a shock to the simplistic view. Not only is \Th exotic but its width
is unprecedentedly small for a strongly decaying resonance.

Lacking in the traditional picture is Quantum Field Theory saying that all baryons
are in fact quantum-mechanical superpositions of $3Q,5Q,7Q,\ldots$ states, and
Spontaneous Breaking of Chiral Symmetry, saying that constituent quarks {\em have}
to interact strongly at least with the pseudoscalar meson fields. A very rough, very
approximate model of baryons that accommodates, however, this physics is the Skyrme model.
There is a prejudice that the Skyrme model is somehow opposite or perpendicular
to quark models. In fact the Skyrme model is in accordance with quarks~\cite{DP-08} but
it is in one way better as it allows baryons to be made of $N_c$ quarks, plus an
indefinite number of $Q\bar Q$ pairs.

The question whether \Th exists in the Skyrme model has been studied, at large $N_c$,
in Ref.~\cite{Klebanov} in the Callan--Klebanov approach, with the conclusion that there
are ``no $S\!=\!+1$ kaon bound states or resonances in the spectrum''. This is
a misunderstanding: the $KN$ cross section in the $T\!=\!0,L\!=\!1$ wave,
computed from the phase shifts in the Skyrme model~\cite{Klebanov} exhibits a strong
resonance around 1500 MeV, see Fig.~1 (the existing data for the full cross section are
also shown there). A more precise way to formulate it is that the $KN$ scattering
amplitude has a pole at $m_{\rm res}-i\frac{\Gamma}{2}=1449-i\,44\,{\rm MeV}$~\cite{DP-08},
for the standard
\vskip -.5true cm

\begin{figure}[htb]
\begin{minipage}[t]{.53\textwidth}
\hspace{-0.48cm}\includegraphics[width=\textwidth]{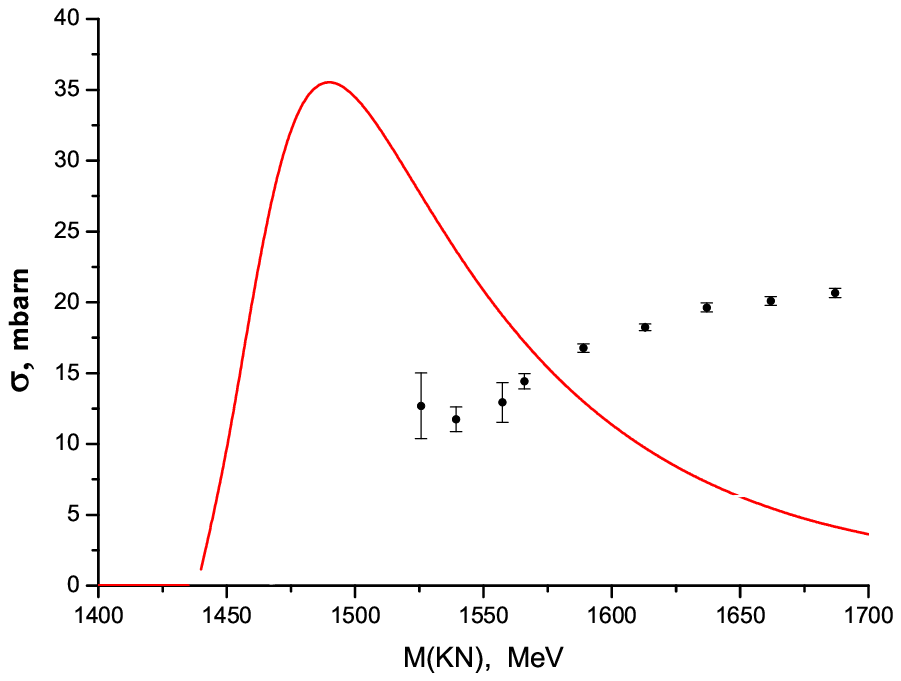}
\vskip -.5true cm

\caption{Skyrme model predicts a strong $\Theta$ resonance.}
\end{minipage}
\hspace{0.2cm}
\begin{minipage}[t]{.44\textwidth}
\vskip -4true cm

\includegraphics[width=\textwidth]{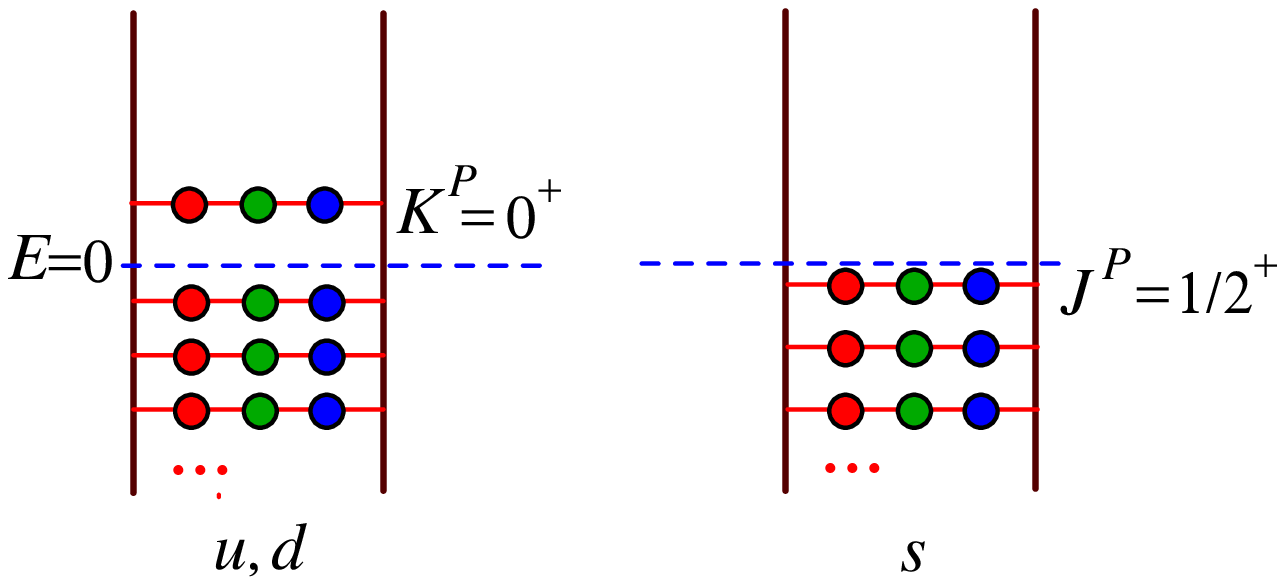}
\vskip 0.81true cm

\caption{Ground-state baryon $N\left(940,1/2^+\right)$.}
\end{minipage}
\end{figure}
\noi
parameters of the Skyrme model. Therefore, the Skyrme model {\em does}
predict a strong exotic resonance, and there is no way to get rid of it as its origin
is very general. However, as explained in detail in Ref.~\cite{DP-08},
the Skyrme model grossly overestimates the \Th width. In realistic settings it becomes
very narrow, and that is why it is so hard to detect it.

There is a nice way to understand \Th in simple terms and see that it is unavoidable~\cite{D-08}.
If the number of quark colors $N_c\!=\!3$ is considered to be a large number, the $N_c$ quarks
in a baryon can be viewed, according to Witten, as bound by a mean field. Any reasonable
{\it Ansatz} for the mean field breaks symmetry between $u,d$ quarks on the one hand and $s$ quarks
on the other. The (approximate) $SU(3)$ symmetry is restored when one considers rotations of
the mean field in flavor space: that produces $SU(3)$ baryon multiplets. The splittings between
multiplet centers are ${\cal O}(1/N_c)$, and the splittings inside multiplets are ${\cal O}(m_s)$.
The ground-state baryon -- the nucleon -- is obtained by filling in all negative-energy one-particle
states in the mean field, and adding one filled shell with positive energy for $u,d$ quarks, see
Fig.~2. The $s$-quark shells are characterized by $J^P$ whereas the $u,d$-quark shells are
characterized by $K^P$ where ${\bf K}\!=\!{\bf J}\!+\!{\bf T}$. The highest shell filled by
$N_c$ quarks in antisymmetric state in color must have $J^P=1/2^+$ for $s$ quarks, and $K^P=0^+$
for $u,d$ quarks. The lowest baryon multiplets $({\bf 8},1/2^+)$ and $({\bf 10},3/2^+)$
are obtained from quantizing the rotations of this filling scheme in flavor and ordinary spaces.

The lowest baryon resonances that are not rotational excitations of the ground-state baryon
are $\Lambda(1405,1/2^-)$ and $N(1440,1/2^+)$. They can be obtained as one-quark excitations
in the mean field: $\Lambda(1405)$ as the $1/2^-$ excitation of the $s$ quark (Fig.~3)
and $N(1440)$ as a $0^+$ (or $1^+$) excitation of $u,d$ quarks (Fig.~4). The existence of the
excited $1/2^-$ level for $s$ quarks implies that it can be also excited by $s$ quark from the highest
filled $s$-quark shell: such particle-hole excitation can be identified with the $N(1535,1/2^-)$
resonance, see Fig.~3. At $N_c\!=\!3$ it is mainly a pentaquark baryon $u(d)uds\bar s$, which
explains its large decay branching into $\eta N$, a long-time mystery.

\begin{figure}[htb]
\begin{minipage}[t]{.48\textwidth}
\includegraphics[width=\textwidth]{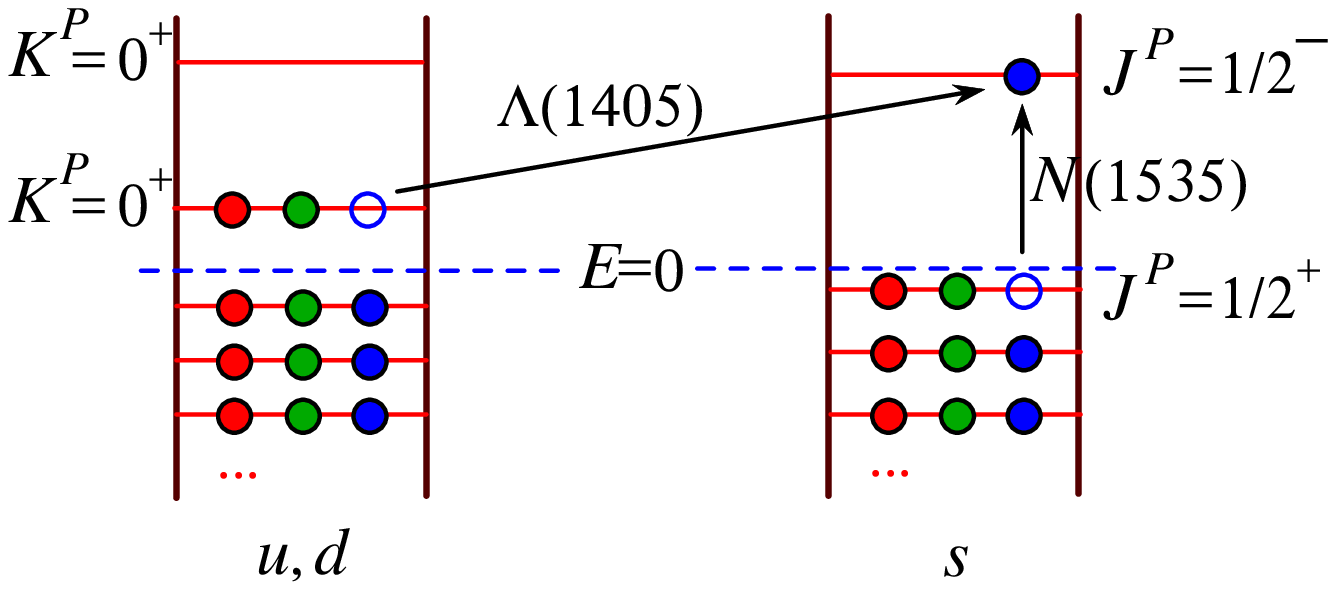}
\caption{One-quark excitations corresponding to $\Lambda(1405,1/2^-)$ and $N(1535,1/2^-)$.}
\label{fig:4}
\end{minipage}
\hspace{0.4cm}
\begin{minipage}[t]{.48\textwidth}
\includegraphics[width=\textwidth]{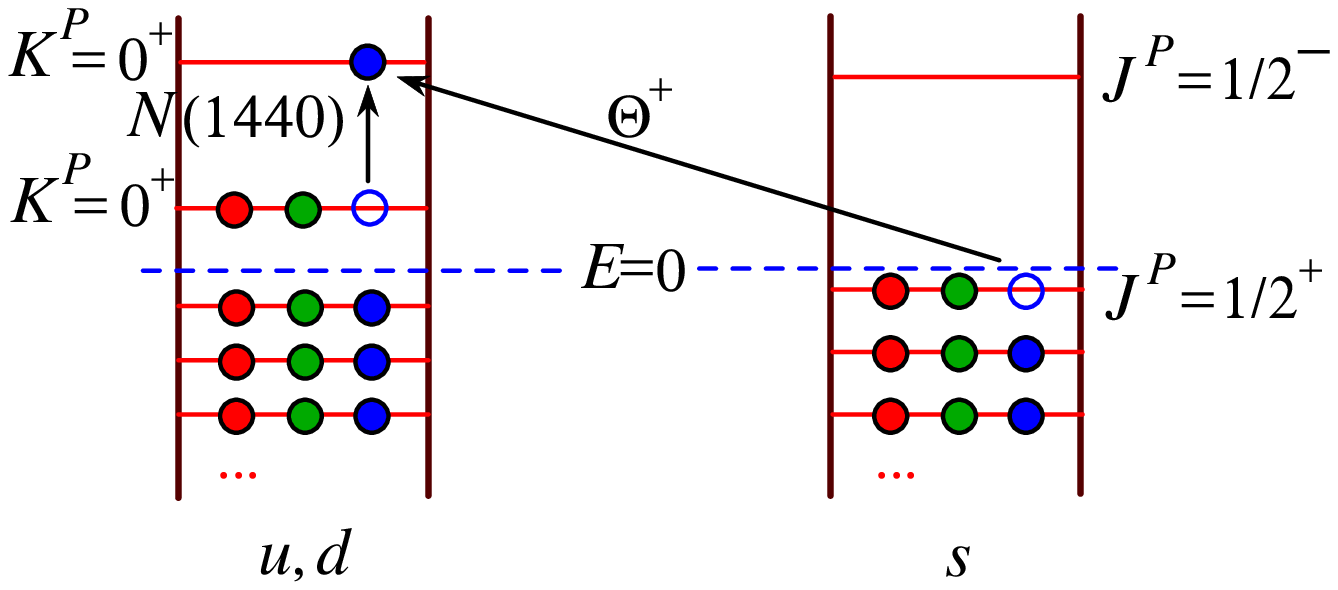}
\caption{One-quark excitations corresponding to $N(1440,1/2^+)$ and $\Theta^+(1/2^+)$.}
\label{fig:5}
\end{minipage}
\end{figure}

The existence of the excited (Roper) $0^+$ level for $u,d$ quarks implies that it can be also
excited from the $s$-quark shell, forming an exotic pentaquark $\Theta^+=uudd\bar s$, see Fig.~4.
Since the positions of all levels involved are already fixed from the masses of the three resonances,
we estimate the \Th mass as $m_{\Theta}\!\approx\! 1440\!+\!1535\!-\!1405\!=\!1570\,{\rm MeV}$,
with the uncertainty of few tens MeV as the resonance masses are not precisely known.
Thus, in the mean field picture the exotic pentaquark is a consequence of the three well-known
resonances and is light~\cite{D-08}. One should not expect a qualitative change in the
levels as one goes from large $N_c$, where the mean field is exact, to the real-world $N_c\!=\!3$.

In this interpretation, \Th is a Gamov--Teller-type resonance long known in nuclear physics,
where a neutron from a filled shell can be put on an excited proton level.
All four excitations shown in Figs.~3,4 entail their own rotational bands of $SU(3)$ multiplets.

Finally, I would like to remark that with all couplings of \Th to ordinary baryons being small,
a promising way of detecting it is in interference with the production of known resonances, since
the interference cross sections are linear (and not quadratic) in the small coupling constants~\cite{Amarian}.

I acknowledge many helpful discussions with Victor Petrov and Klaus Goeke.

\end{document}